\documentclass[twocolumn]{aastex631}

\usepackage{url}

\usepackage{amsmath,bm}
\usepackage{float}
\usepackage{rotating}
\usepackage{gensymb}
\usepackage{hyperref}
\usepackage{xcolor}
\usepackage{multirow}
\graphicspath{{./}{Figures/}}
\usepackage{graphicx}  
\usepackage{booktabs}
\shorttitle{}
\shortauthors{}
\usepackage{orcidlink}
\usepackage{cancel}

\begin{document}

\title{Eccentricity Evolution of Warm Jupiters: The Role of Distant Perturbers and Nearby Companions}
\author{Ying He}
\affiliation{Department of Physics, Anhui Normal University, Wuhu Anhui, 241000, People's Republic of China}
\author[0000-0001-9424-3721]{Dong-Hong Wu}
\affiliation{Department of Physics, Anhui Normal University, Wuhu Anhui, 241000, People's Republic of China}
\author[0000-0002-9063-5987]{Sheng Jin}
\affiliation{Department of Physics, Anhui Normal University, Wuhu Anhui, 241000, People's Republic of China}

\correspondingauthor{Dong-Hong Wu}
\email{wudonghong@ahnu.edu.cn}

\begin{abstract}
Warm Jupiters---giant exoplanets with orbital periods between 10 and 200 days---exhibit a broad range of eccentricities and are often accompanied by nearby low-mass planets. Understanding the origins of their orbital architectures requires examining both their migration histories and subsequent dynamical interactions. In this study, we perform extensive $N$-body simulations to explore how distant giant planet perturbers affect the eccentricity evolution of warm Jupiters and the role of nearby super-Earth companions in mediating these interactions. We find that while distant perturbers can induce large-amplitude eccentricity oscillations in warm Jupiters via the von Zeipel–Lidov–Kozai mechanism, the presence of nearby super-Earth companions often suppresses these variations via strong dynamical coupling. This mechanism naturally leads to a bimodal eccentricity distribution: warm Jupiters with nearby companions tend to maintain low eccentricities, whereas those without exhibit significantly broader eccentricity distributions. We show that reproducing the observed eccentricity distribution of warm Jupiters lacking nearby companions is most naturally explained if a substantial fraction of distant perturbers occupy dynamically extreme orbits, either with large mutual inclinations or high orbital eccentricities. These results support a scenario in which warm Jupiters experience substantial post-disk dynamical evolution, shaped jointly by distant perturbers and nearby companions.

\end{abstract}


\section{Introduction} 
\label{section:intro}

Warm Jupiters---giant planets with orbital periods of 10-200 days and masses $M \sin i > 0.25 \ M_{\rm {Jup}}$---occupy a distinctive ``valley'' in the period distribution of exoplanets \citep{Dawson2018}. Despite their theoretical significance as potential intermediaries between hot Jupiters ($P <$ 10 days) and long-period giants ($P>200$ days), only about 200 warm Jupiters have been detected to date, far fewer than their shorter- and longer-period counterparts. Like other close-in giants, they are thought to have formed in the outer disk and migrated inward; however, the precise origin of warm Jupiters remains debated. Two primary mechanisms have been proposed: (1) high-eccentricity tidal migration, driven by perturbations from distant giant planets or stellar companions, followed by tidal dissipation  \citep{Wu_2003,Petrovich2015KL,Petrovich2016,Hamers2017ApJ}; and (2) quiescent disk migration, mediated by interactions with the protoplanetary disk \citep[e.g.,][]{Lin1986,Ida2008}. The former scenario often casts warm Jupiters as progenitors of hot Jupiters, while the latter implies smoother orbital evolution. 

However, both models face significant challenges. For high-eccentricity migration to occur, warm Jupiters must attain extreme eccentricities ($e\geq0.9$) to permit tidal decay,  yet most observed warm Jupiters exhibit modest eccentricities $(<0.9)$ and orbital distances where tidal effects are weak ($a(1-e^2)>0.1$ au). One proposed resolution is that these planets undergo eccentricity oscillations. Current observations may capture them in a low-$e$ phase. The von Zeipel–Lidov–Kozai (vZLK)oscillation has been invoked to explain such dynamics \citep{Wu_2003,Fabrycky2007,Petrovich2015KL,Anderson2016MNRAS}, but it requires a relatively massive or close-in companion to overcome general relativistic suppression \citep{Dong_2014}. If the perturber is too distant, the eccentricity of the migrating planet would be frozen to high values, leading to rapid orbital decay and the eventual formation of a hot Jupiter instead of a warm Jupiter \citep{Petrovich2016,Antonini2016,Vick2019}. 

Several observational trends further complicate the high-eccentricity scenario. First, population synthesis models predict that high-eccentricity tidal migration produces far fewer warm Jupiters than are actually observed \citep{Petrovich2016,Antonini2016,Hamers2017MNRAS,Vick2019}. Second, over $50\%$ of warm Jupiters are observed to host nearby low-mass companions \citep{Huang2016,Wu2023(70)}, and these systems typically have low eccentricities---difficult to reconcile with violent migration (though exceptions exist, such as KOI-984 c, which has an orbital eccentricity of 0.4 and a nearby companion \citep{Sun2022}). Finally, Rossiter-McLaughlin measurements reveal that warm Jupiters tend to exhibit low stellar obliquities, remaining aligned with their host star's spin axes \citep{Wang2021,Rice2022,Wright2023,Lubin2023,Wang2024}, inconsistent with the chaotic dynamics expected from high eccentricity migration.

Disk migration, meanwhile, struggles to explain the broad eccentricity distribution of warm Jupiters \citep{Gupta2023,Dong2023,Dong2025,Bieryla2025,Morgan2025}, particularly those with extremely high eccentricities ($e>0.9$), such as HD 80606 b \citep{Naef2001} and TIC 241249530 b \citep{Gupta2024}, since disk-planet interactions typically dampen eccentricity \citep{Goldreich2003,Kley2012}. Recent studies have highlighted the role of post-disk dynamical evolution in shaping the architecture of warm Jupiter systems \citep{Wu2023,He2024}. Mechanisms such as secular perturbations from massive outer companions \citep{Anderson2017} and \emph{in situ} scattering among planets \citep{Anderson2020} may broaden the eccentricity distribution of warm Jupiters.

Based on the analytical framework developed by \citet{Anderson2017}, who proposed that nearby low-mass planets can shield warm Jupiters from the vZLK excitation, we performed large-scale $N$-body simulations to systematically quantify this effect across a population of planetary systems. In this work, we consider scenarios in which warm Jupiters form with initially low eccentricities and host nearby super-Earths. These systems then undergo eccentricity oscillations driven by secular gravitational perturbations from distant giant planets. We investigate how outer companions influence the dynamical evolution of inner warm Jupiter systems and assess the role of nearby super-Earths in modulating the eccentricity oscillations. 

The paper is organized as follows. Section \ref{sec:simulation setup} outlines the setup of our simulations. In Section \ref{sec:results}, we present our results, focusing on the mechanisms driving or suppressing eccentricity excitation in warm Jupiters and the resulting eccentricity distributions. In Section \ref{sec:observation}, we compare our simulated results with observations. Finally, we summarize our findings and draw conclusions in Section \ref{sec:conclusion}.

\section{Simulation setups}\label{sec:simulation setup}

We simulated a set of planetary systems, each consisting of three planets: a warm Jupiter ($m_{\rm WJ}=1\ M_{\rm {Jup}}$), a closely accompanying super-Earth with mass $m_{\rm SE}$, and a distant giant perturber with mass $m_{\rm {per}}$, all orbiting a solar-mass star ($1\ M_{\odot}$).  The orbital period of the warm Jupiter was randomly distributed between 10 and 200 days. The mass of the super-Earth, $m_{\rm SE}$, was drawn from a normal distribution with a mean of 6 $M_{\oplus}$ and a standard deviation of 1 $M_{\oplus}$, constrained to $m_{\rm SE}>3 $ $M_{\oplus}$. The period ratio between the super-Earth and the warm Jupiter was sampled from a uniform distribution ranging from 1.4 to 4, allowing the super-Earth to orbit either inside or outside the orbit the warm Jupiter. The lower limit ensures the classical two-planet Hill stability, as period ratios $<1.4$ typically correspond to separations less than 3.5 mutual Hill radii---below the stability threshold for two-planet systems. 

The semi-major axis of the distant perturber, $a_{\rm per}$, was uniformly distributed between 1 and 10 au, and its mass, $m_{\rm per}$, was uniformly distributed between 1 and 5 $M_{\rm {Jup}}$. The perturber's orbital inclination followed a normal distribution centered at $50^{\circ}$ with a $20^{\circ}$ standard deviation, allowing us to study inclined configurations. This choice was designed to ensure that a significant fraction of systems reside within the high inclination regime ($i \gtrsim 40^\circ$) where the vZLK mechanism is effective, thereby facilitating a clear examination of eccentricity excitation and its potential suppression by nearby super-Earth. All planets began with eccentricities drawn from a Rayleigh distribution with scale parameter ${\sigma=0.04}$ (i.e., ${ e}\sim\mathrm R(0.04)$), while the angular orbital elements (argument of pericenter, longitude of ascending node, and initial mean anomaly) were each uniformly distributed between $0$ and 360$^\circ$. The inner planets' initial inclinations were set to approximately half their eccentricities. 

\begin{figure*}[htbp]
  \centering
   {
       \begin{minipage}[h]{0.45\linewidth}
       \includegraphics[width=\linewidth]{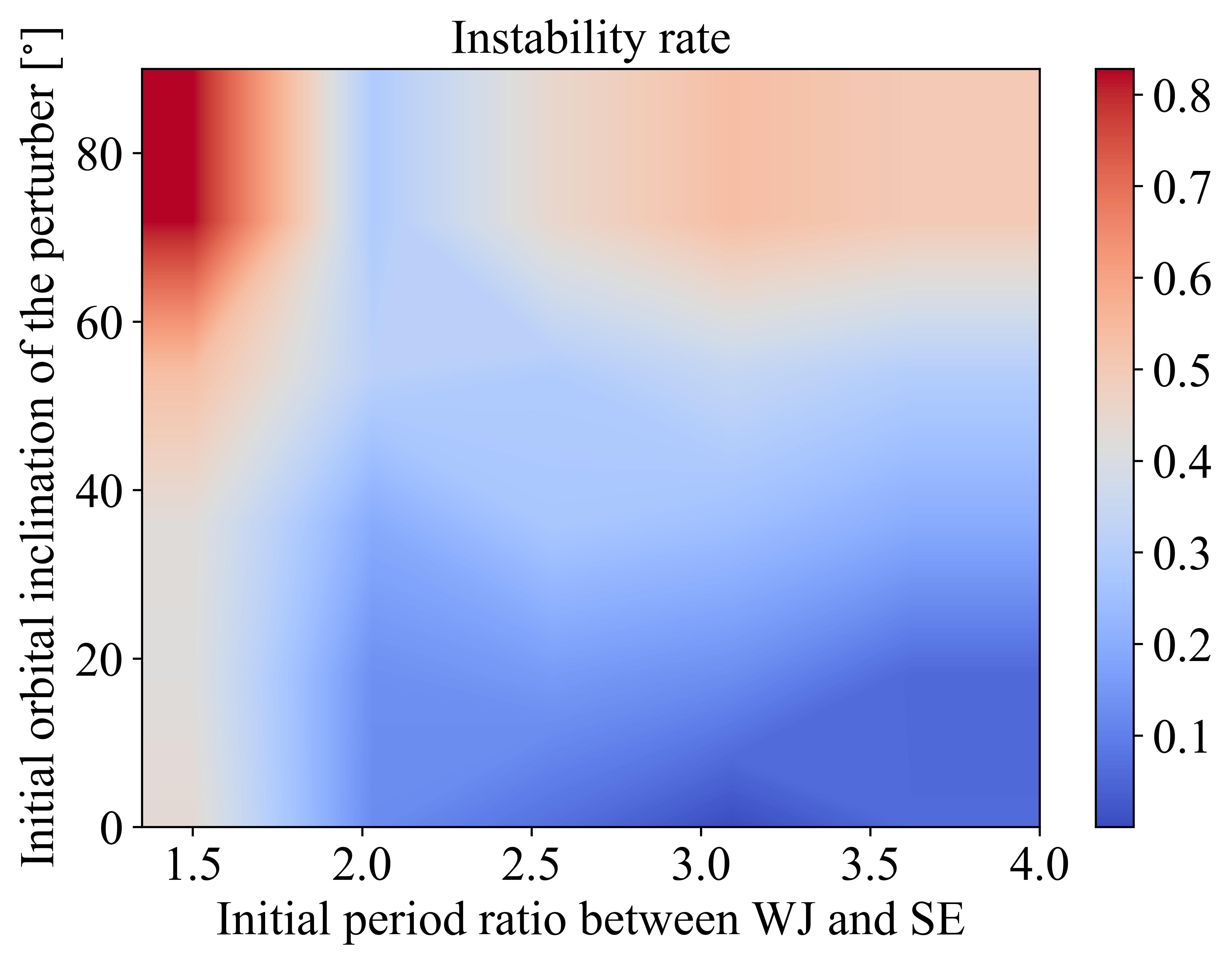}
       \end{minipage}
    }
   {
       \begin{minipage}[h]{0.45\linewidth}
       \includegraphics[width=\linewidth]{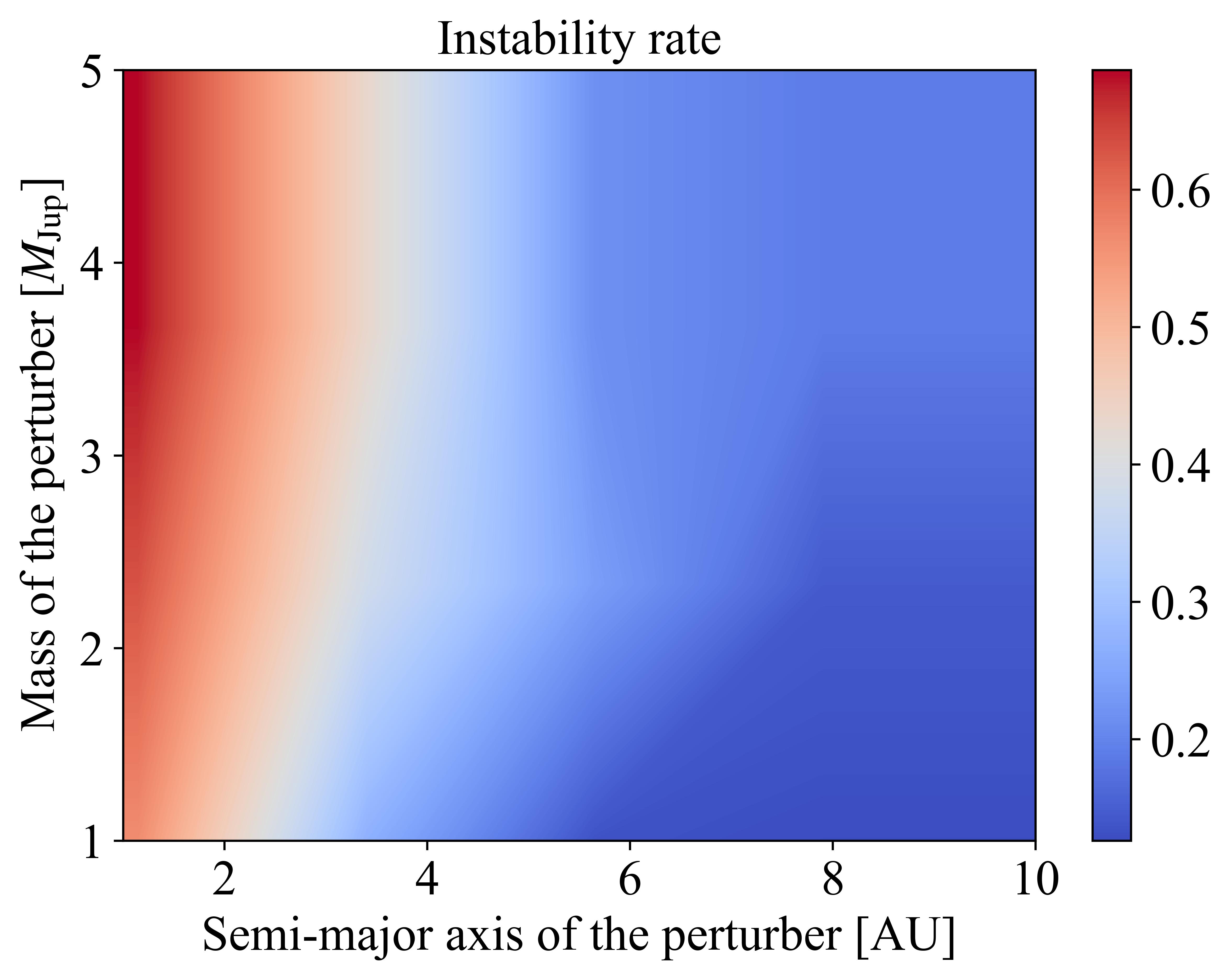}
       \end{minipage}
    }
    \caption{The instability rates as functions of the initial warm Jupiter (WJ)-super-Earth (SE) period ratio and perturber properties. The left panel shows how system stability depends on the inner pair's period ratio and the perturber's inclination, while the right panel shows the impact of the perturber's mass and orbital distance on dynamical instability. }
  \label{fig:Instability rate}
\end{figure*}

We generated 2000 such systems for our fiducial simulations. All systems were integrated using the \texttt{MERCURIUS} hybrid integrator within the REBOUND package \citep{Rein2012,Rein2019}, with general relativistic precession effects included via the \texttt{gr-potential} option in the REBOUNDx package \citep{Tamayo2020}. In our simulations, planetary collisions were allowed. The radii of super-Earths were calculated using the relation $R/R_{\oplus} = (M/M_{\oplus})^{0.59}$ \citep{Chen2017}, while giant planets were assigned a fixed radius of 11.2 $R_{\oplus}$. Collisions were triggered when the separation between two bodies fell below the sum of their radii, resulting in a merger that conserved total mass and angular momentum. Planets with semi-major axis exceeding 100 au from the host star were considered ejected from the system and excluded from further simulation. The integration was terminated either when the system reached a maximum duration of $10^7$ yr or when only one planet remained. Note that tidal dissipation effects were neglected in our simulations, as only a small fraction of warm Jupiters in our sample reached the tidal migration threshold.

\section{results and discussion}\label{sec:results}
\subsection{Dynamical Instability}

Distant perturbers can significantly influence the orbital characteristics of inner systems. Previous studies \citep{Pu2018} have shown that such perturbers are capable of exciting both the eccentricities and mutual inclinations of inner planets. The dynamical influence is particularly pronounced for perturbers with high values of $m_{\rm per}/a_{\rm per}^3$ and substantial orbital misalignment relative to the inner system. 

We investigated how the dynamical stability of planetary systems depends on four key parameters (Figure \ref{fig:Instability rate}):  (1) the initial period ratio between the warm Jupiter and the super-Earth, (2) the perturber's inclination, (3) its mass, and (4) its semi-major axis. Systems are classified as unstable if they fail to retain all three planets over a $10^7$-yr integration.  Our results indicate that stability is generally enhanced when the inner planet pair's period ratio exceeds 1.7 and the outer perturber is well aligned with the inner system. Additionally, perturbers located beyond 4 au exert minimal destabilizing effects on warm Jupiters. 

Instability typically emerges through orbital crossings and close encounters, ultimately leading to either collisions or ejections. Our simulations reveal that planet loss most commonly occurs via ejection ($76.8 \%$ for the super-Earth and $11.8\%$ for the warm Jupiter), followed by planet-planet collisions ($20.1 \%$). 

Notably, systems with an inner pair's period ratio between 1.7 and 2.5 exhibit greater stability than those with wider spacings---particularly when the perturber's inclination exceeds $60^\circ$. This stability enhancement likely stems from stronger dynamical coupling between inner planets at smaller period ratios, which suppresses mutual inclination growth and helps preserve the integrity of the system's architecture.

\begin{figure}[h]
  \centering
   \includegraphics[width=0.45\textwidth]{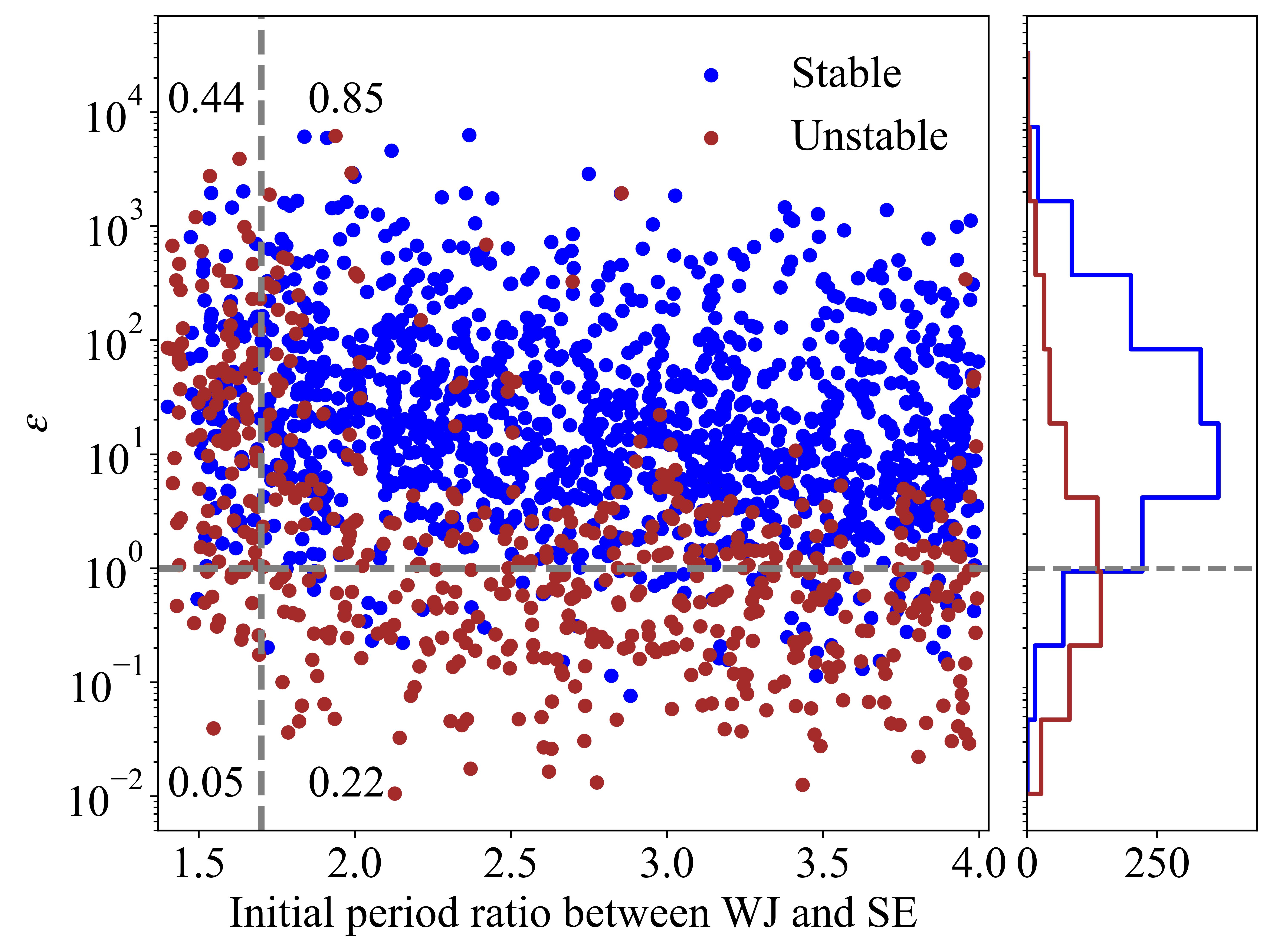}
    \caption{\textbf{Left}: distribution of the coupling parameter $\varepsilon$ and the inner pair's period ratio for stable (blue dots) and unstable (dark red dots) systems. The gray dashed lines mark $\varepsilon=1$ and a period ratio of 1.7. Numbers in each region indicate the corresponding stability rates. \textbf{Right}: distribution of $\varepsilon$ for stable (blue) and unstable (dark red) systems with inner period ratios $>1.7$. }
  \label{fig:eps-period ratio}
\end{figure}

The coupling strength between the warm Jupiter and the super-Earth is quantified by the parameter $\varepsilon$, as defined by \citet{Anderson2017}, which represents the ratio of the precession rate induced by the nearby super-Earth to that induced by the distant perturber:
\begin{equation}
\varepsilon = 
\begin{cases} 
  \frac{3m_{\rm SE}}{m_{\rm per}} \frac{a_{\rm per}^3}{a_{\rm SE}^3}, & \text{if } a_{\rm SE} > a_{\rm WJ} \\
  \frac{3m_{\rm SE}}{m_{\rm per}} \frac{a_{\rm SE}^2 a_{\rm per}^3}{a_{\rm WJ}^5}, & \text{if } a_{\rm SE}<a_{\rm WJ}
\end{cases}
\end{equation} where only the first-order terms in the expansion of either $a_{\rm WJ}/a_{\rm SE}$ or $a_{\rm SE}/a_{\rm WJ}$  in the Laplace coefficient have been retained. When $\varepsilon >1$, the super-Earth dominates the precession, suppressing eccentricity excitation in the warm Jupiter. In contrast, $\varepsilon <1$ indicates perturber-dominated dynamics, resulting in substantial eccentricity excitation. 

As shown in Figure \ref{fig:eps-period ratio}, $\varepsilon$ is anti-correlated with the inner pair's period ratio, with a Spearman Rank correlation coefficient of -0.20 and a $p$-value $<10^{-4}$. Systems with inner pair period ratios $<2.5$ exhibit stronger coupling, with 87.0\% having $\varepsilon >1$ compared to 79.0\% for systems with larger period ratios. The stability implications are particularly striking---stable systems generally exhibit significantly higher $\varepsilon$ values than unstable ones, as illustrated in the right panel of Figure \ref{fig:eps-period ratio}. Our results show that systems with $\varepsilon>1$ exhibit stability rates that are nearly 4 to 9 times that of their weakly coupled ($\varepsilon<1$) counterparts. In particular, systems with $\varepsilon>1$ and inner period ratios $>1.7$ show a stability rate of 0.85, which is 17 times that of systems with $\varepsilon<1$ and inner period ratios $<1.7$. These results suggest that strong coupling between the inner planets plays a crucial role in enhancing overall system stability. This is consistent with the secular stability criteria given by previous studies \citet{Denham2019} and \citet{Wei2021}, which propose that strong internal precession suppresses eccentricity excitation from a distant perturber.

For systems in which the perturber's inclination exceeds $40^\circ$, the vZLK mechanism \citep{Lidov1962,Kozai1962} can be triggered, driving periodic oscillations in the eccentricities and inclinations of both the warm Jupiter and the super-Earth. If the inner pair is weakly coupled, their mutual inclination can be excited to large values \citep{Lai2017}, potentially leading to dynamical instability. Among systems with inner pair period ratios between 1.7 and 2.5, approximately $72.6\%$ maintain mutual inclinations below $5^\circ$, compared to $58.9\%$ for systems with period ratios $>2.5$. This trend helps explain why, in systems with highly inclined perturbers (inclination $>60^\circ$), dynamical stability is enhanced when the inner period ratio lies between 1.7 and 2.5.

When the perturber's orbital inclination is below  $40^\circ$, the orbital inclinations of the inner planets are not significantly excited, and their mutual inclination typically remains below $5^{\circ}$. In this regime, about $76.0\%$ of systems with inner pair period ratios between 1.7 and 2.5, and $82.1\%$ of systems with inner pair period ratios $>$ 2.5, exhibit mutual inclinations below $5^{\circ}$.

Additionally, closer and more massive perturbers correspond to smaller values of $\varepsilon$, resulting in weaker coupling between the inner planets and making them more prone to instability, as illustrated in the right panel of Figure \ref{fig:Instability rate}.

\subsection{Eccentricity Distribution}

\subsubsection{Fiducial Simulation Results}\label{ecc_fid}

\begin{figure}
  \centering
   \includegraphics[width=0.45\textwidth]{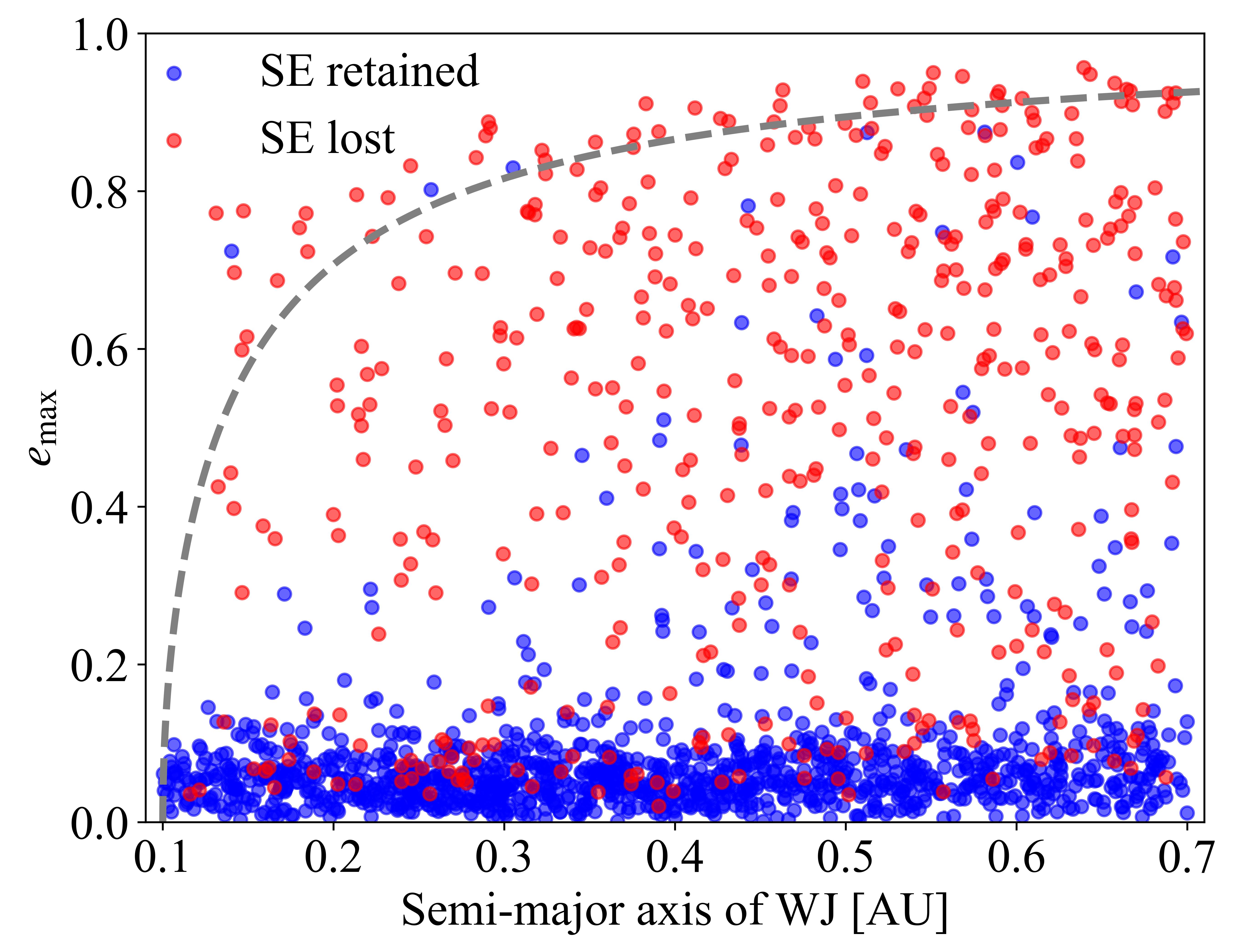}
    \caption{Maximum eccentricity versus semi-major axis of warm Jupiters in our simulations. Blue dots represent warm Jupiters that retained nearby super-Earths, while red dots indicate warm Jupiters that lose their nearby super-Earths. The dashed curve marks the tidal migration threshold, defined by $R_{\rm p} =a(1-e^2)=$ 0.1 au.}
  \label{fig:a_e}
\end{figure}

In our simulations, the quadrupole-level vZLK timescales ($t_{\rm vZLK}$) for the warm Jupiters---calculated using Equation (1) from \citet{Petrovich2016}---range from $10^3$ to $10^6$ yr, with all systems satisfying $t_{\rm vZLK}<10^7$ yr. Therefore, if the vZLK mechanism is activated, the recorded maximum eccentricity ($e_{\rm max}$) represents the true peak the warm Jupiter can attain during its dynamical evolution. A detailed assessment of octupole‑level contributions and the sufficiency of the integration time is provided in Section \ref{quad_oct}. The relationship between the maximum eccentricity and semi-major axis of the simulated warm Jupiters is shown in Figure \ref{fig:a_e}. For reference, the tidal migration threshold, defined by $R_{\rm p}=a(1-e^2)= 0.1$ au, is indicated by a dashed curve. Planetary systems falling below this threshold are unlikely to undergo significant tidal migration. While some warm Jupiters attain high eccentricities, the majority remain relatively circular ($e<0.2$), with only 2.6\% reaching eccentricities high enough to cross the  tidal migration boundary. Notably, warm Jupiters with nearby super-Earth companions tend to exhibit the lowest eccentricities, likely due to the stabilizing influence of strong dynamical coupling with their nearby companions. 

As established by \citet{Anderson2017}, eccentricity excitation of the inner planet occurs when inclination of the distant perturber lies within the vZLK window. Outside this window, warm Jupiters maintain relatively low eccentricities ($e < 0.2$). Once the orbital configuration enters the vZLK regime, oscillations are triggered that typically drive up the eccentricity of the warm Jupiter. However, strong gravitational coupling between the warm Jupiter and its nearby super-Earth companion can suppress such excitation.  In the following section, we examine in greater detail how the presence of nearby super-Earths influences the eccentricity evolution of warm Jupiters. 

\subsubsection{Influence of Nearby super-Earth}\label{sec:Nearby super-Earth}

\begin{figure}[htbp]
  \centering
   \includegraphics[width=0.45\textwidth]{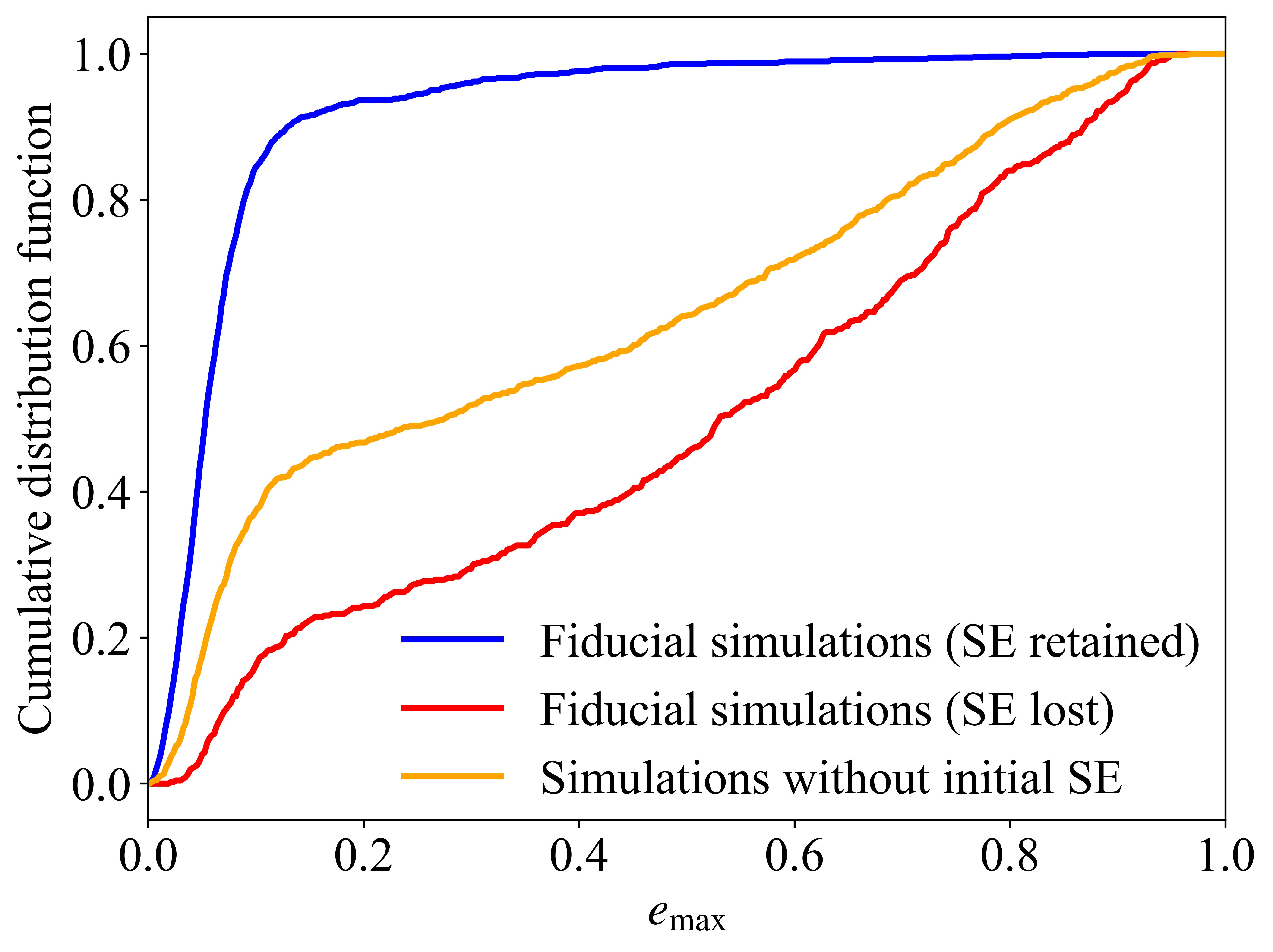}
\caption{Cumulative distribution function of the maximum eccentricity reached by warm Jupiters. Results are shown for: warm Jupiters that retain nearby super-Earths from our fiducial simulations (blue), warm Jupiters that lose nearby super-Earths (red), and warm Jupiters that initially lack nearby super-Earths (orange).}
  \label{fig:e(SE vs noSE)}
\end{figure}

To assess the influence of nearby super-Earths on the eccentricity excitation of warm Jupiters, we performed an additional set of 1000 simulations. The initial conditions and parameter distributions were identical to those used in our fiducial setup described in Section \ref{sec:simulation setup}, except that super-Earths were excluded. The resulting maximum eccentricity distributions of warm Jupiters in systems with and without nearby super-Earth companions are compared in Figure \ref{fig:e(SE vs noSE)}.

We find that warm Jupiters without nearby super-Earths---either initially absent or lost due to dynamical interactions---exhibit significantly higher eccentricity excitation than those that retain super-Earth companions. This behavior can be primarily attributed to the damping effect that super-Earths exert on the eccentricity of warm Jupiters through gravitational coupling. In stable systems where super-Earths persist, the coupling parameter $\varepsilon$ typically exceeds unity (Figure \ref{fig:eps-period ratio}), enabling the super-Earth to dominate the orbital precession of warm Jupiters. This strong coupling suppresses the vZLK-induced eccentricity growth of warm Jupiters driven by outer perturbers.

\begin{figure}[htbp]
  \centering
   \includegraphics[width=0.45\textwidth]{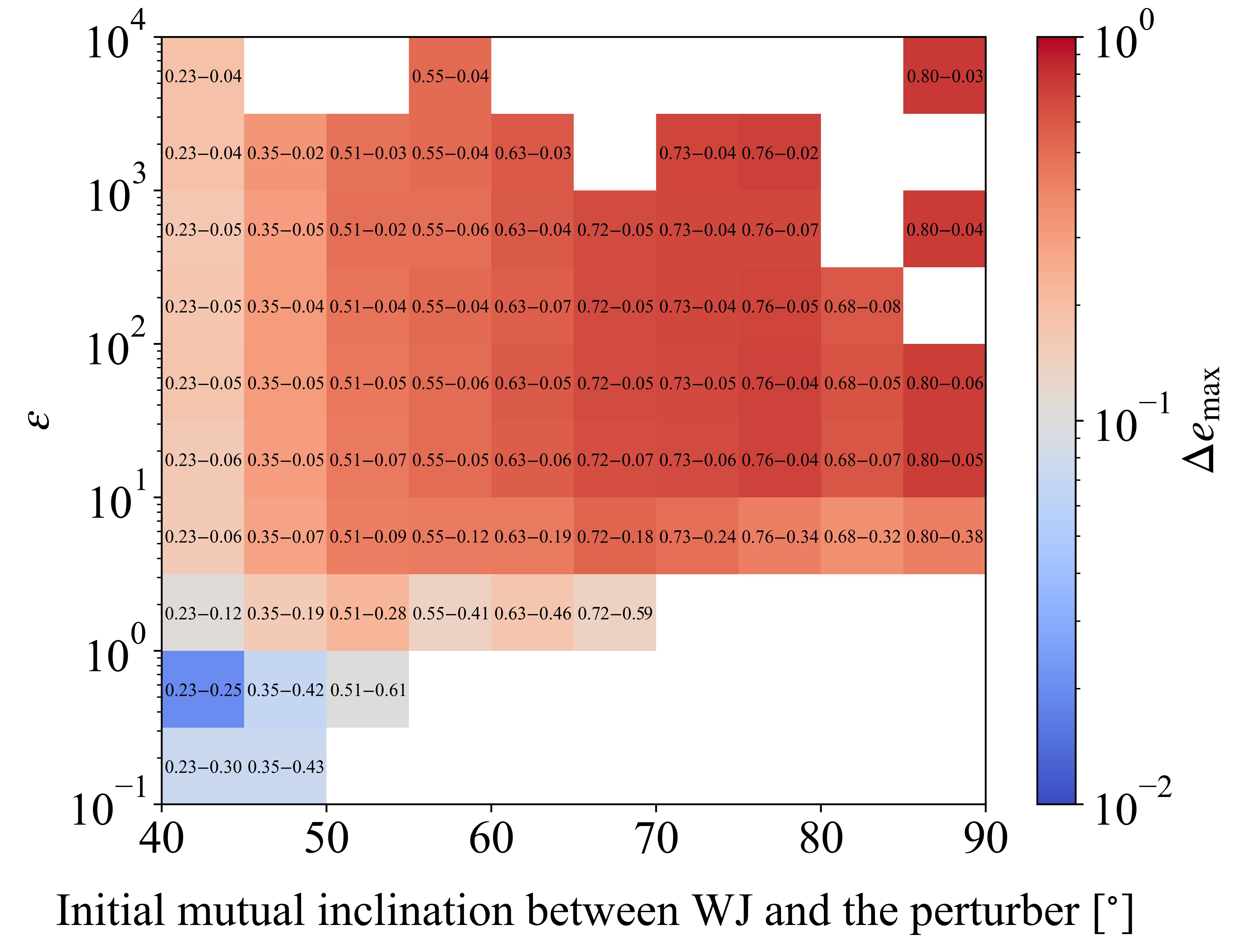}
    \caption{$\Delta e_{\rm max}$ across the $\varepsilon$---mutual inclination space. Colors denote the absolute difference in mean $e_{\rm max}$ between warm Jupiters without primordial super-Earths and those retaining them. The first value in each cell represents the mean $e_{\rm max}$ of warm Jupiters without primordial super-Earths, while the second value corresponds to systems that consistently host nearby super-Earths. Blank cells indicate no stable warm Jupiters retaining nearby super-Earths in the bin.}
  \label{fig:Delta_emax}
\end{figure}

To quantify this effect, we compare the absolute difference in maximum eccentricity  between warm Jupiters that initially lacked nearby super-Earths and those that consistently retained them throughout the simulation (Figure \ref{fig:Delta_emax}). For systems with $\varepsilon <1$, the resulting $e_{\rm max}$ of warm Jupiters is similar regardless of the presence of an initial super-Earth. However, in systems with $\varepsilon >1$, warm Jupiters that retained nearby super-Earths exhibit significantly lower $e_{\rm max}$ than their counterparts that lacked such companions. This difference becomes even more pronounced in systems with highly inclined outer perturbers. Furthermore, as $\varepsilon$ increases, the maximum eccentricity of warm Jupiters with stable nearby companions decreases systematically. In particular, when $\varepsilon>10$, $e_{\rm max}$ is generally below 0.1. These findings suggest that the presence of a dynamically stable super-Earth effectively suppresses eccentricity excitation in warm Jupiters, particularly when strong inner coupling is present.

\begin{figure}
    \centering
    \includegraphics[width=0.45\textwidth]{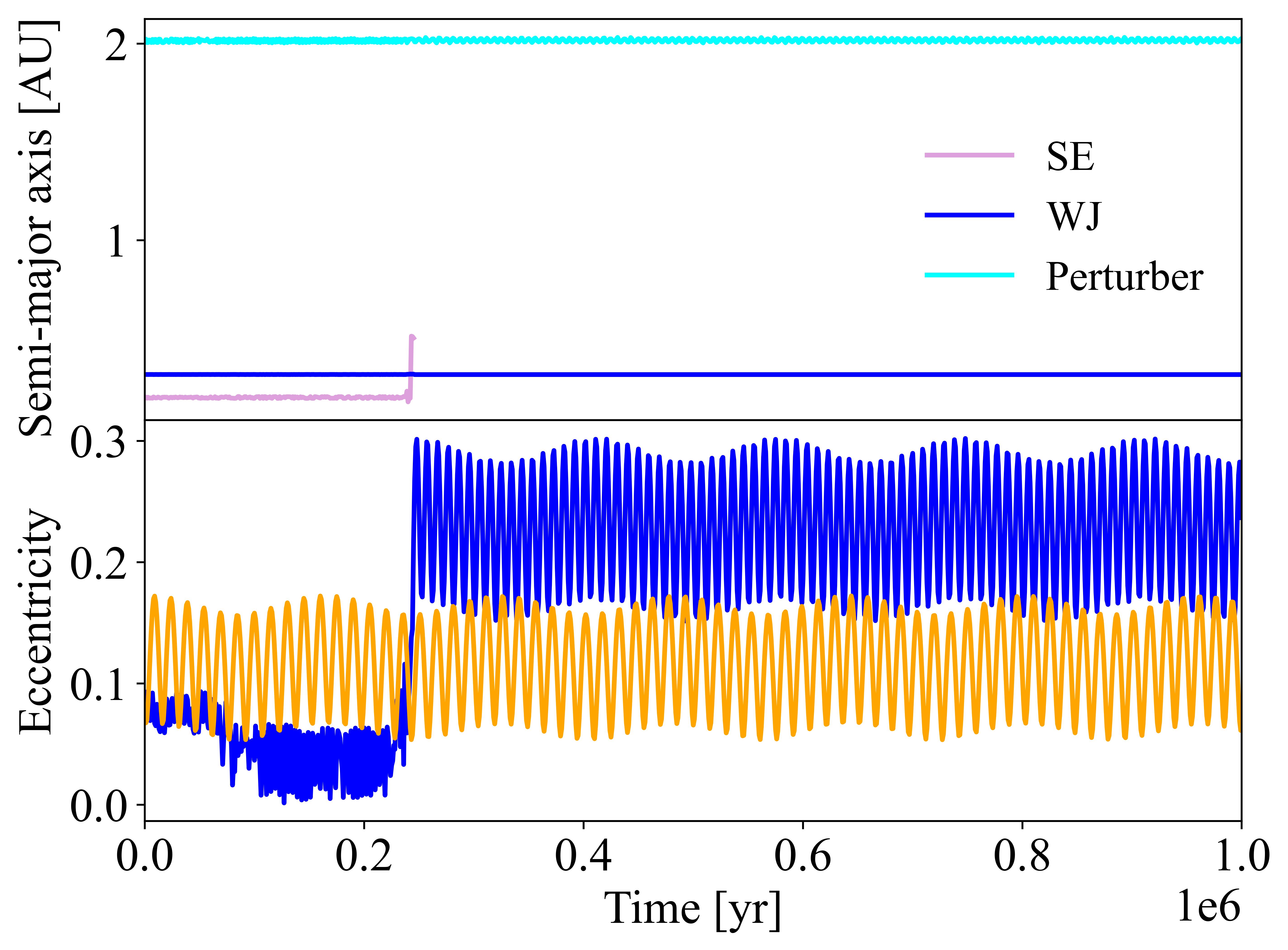}
    \caption{Dynamical evolution over $10^6$ yr of a planetary system consisting of (1) a 6 $M_{\oplus}$ super-Earth on a near-circular orbit ($a$ = 0.2 au, $e$ = 0.04, $i$ = 0.04), (2) a typical warm Jupiter ($a$ = 0.3 au, $e$ = 0.08, $i$ = 0.02), and (3) a distant 2 $M_{\rm Jup}$ perturber ($a$ = 2.0 au, $e$ = 0.06, $i$ = 0.65). The super-Earth, warm Jupiter and perturber are shown in plum, blue and cyan lines, respectively. The super-Earth is ejected at approximately $2.3\times 10^5$ yr. The orange line represents the system evolved without the super-Earth but with otherwise identical initial conditions. }
    \label{fig:SE lost} 
\end{figure}

For warm Jupiters that lose their nearby super-Earths due to dynamical instability, their orbital eccentricities are significantly more excited than those that retain their nearby companions. Once a super-Earth is ejected, the damping effect it exerts on the warm Jupiter disappears, leading to higher orbital eccentricities compared to systems in which the super-Earth remains bound. Figure \ref{fig:SE lost} illustrates the evolution of semi-major axis and eccentricity for two representative systems: one including a super-Earth, and the other without. Aside from the presence of the super-Earth, the two systems share identical initial parameters. Prior to the super-Earth's ejection (at approximately $2.3\times 10^5$ yr), the warm Jupiter in the system with the super-Earth exhibits a lower eccentricity than its counterpart. However, after the ejection, the eccentricity of the warm Jupiter increases rapidly, eventually exceeding that in the system without a super-Earth. This behavior can be attributed in part to the loss of angular momentum associated with the ejection of the super-Earth, which further amplifies the eccentricity of the remaining warm Jupiter. Consequently, eccentricity excitation of warm Jupiters that lose nearby super-Earths is notably greater than that of systems that either retain such companions or never possessed them.

\subsection{External Influences on Eccentricity Evolution}

\subsubsection{Outer perturber mass and eccentricity}

In our fiducial simulations, the initial mass of the outer perturber $m_{\rm per}$ was uniformly distributed between 1 and 5 $M_{\rm Jup}$ (i.e. $m_{\rm per} \thicksim \rm U(1,5)\ M_{\rm Jup}$). To systematically investigate the influence of $m_{\rm per}$ on the eccentricity excitation of warm Jupiters, we conducted three additional simulation sets (1000 systems each), in which $m_{\rm per}$ spanned the ranges of 5–10, 10–15, and 15–20 $M_{\rm Jup}$, respectively. Furthermore, to explore the role of the perturber's initial eccentricity, $e_{\rm per}$, we performed three complementary sets of simulations (500 systems each), in which the initial eccentricity of the outer perturber $e_{\rm per}$ was fixed at 0.1, 0.3, and 0.5, while all other parameters were held at their fiducial values.

\setlength{\tabcolsep}{5pt}
\begin{table}[t]
\centering
\caption{Eccentricity distribution statistics for warm Jupiters in simulations with varying $m_{\rm per}$ and initial $e_{\rm per}$.}
\begin{tabular}{cccc}
    \toprule
    \\[-3.9ex]
    \toprule
    Simulation & $f(e>0.2)$ & $e_{\rm max}$ & $f$($R_{\rm p} <$ 0.1 au) \\ 
    \midrule
    Fiducial simulation & 24.6\% & 0.957 & 2.6\% \\
    $m_{\rm per} \sim \rm{U}(5,10) \ \it{M}_{\rm Jup}$ & 33.7\% & 0.961 & 4.0\% \\
    $m_{\rm per} \sim \rm{U}(10,15)\ \it{M}_{\rm Jup}$ & 36.7\% & 0.968 & 4.4\% \\
    $m_{\rm per} \sim \rm{U}(15,20)\ \it{M}_{\rm Jup}$ & 43.9\% & 0.976 & 6.1\% \\
    $e_{\rm per}$ = 0.1 & 26.9\% & 0.942 & 2.6\% \\
    $e_{\rm per}$ = 0.3 & 27.0\% & 0.949 & 1.4\% \\
    $e_{\rm per}$ = 0.5 & 30.8\% & 0.983 & 5.5\% \\
    \bottomrule
\end{tabular}
\label{Tab:different mass&ecc of perturber}
\end{table}

Table \ref{Tab:different mass&ecc of perturber} summarizes the eccentricity behavior of warm Jupiters across our simulations. As the perturber mass $m_{\rm per}$ increases, the fraction of warm Jupiters with $e > 0.2$ steadily rises---from 24.6\% in the fiducial case to 43.9\% for the most massive perturbers---and the maximum eccentricity increases accordingly, from 0.957 to 0.976. Notably, the fraction of planets surpassing the tidal dissipation threshold ($R_{\rm p}=a(1-e^2) < 0.1$ au) more than doubles from 2.6\% to 6.1\%.

Regarding the initial eccentricity $e_{\rm per}$, increasing it from 0.1 to 0.5 yields a modest rise in the fraction of warm Jupiters with $e > 0.2$, from 26.9\% to 30.8\%, and the maximum eccentricity also increases to 0.983. The potential tidal migration rate similarly increases, from 2.6\% to 5.5\%, though the overall occurrence remains relatively low.

In summary, both the mass and initial eccentricity of the outer perturber significantly influence the eccentricity distribution of warm Jupiters. A more massive perturber would enhance the eccentricity excitation and increase the likelihood of tidal migration for warm Jupiters. Higher initial eccentricity further contributes to this trend, albeit to a lesser degree. Nevertheless, even under favorable conditions, the fraction of systems undergoing tidal migration remains below 7\%, implying that additional factors---such as specific orbital architectures or stellar mass perturbers---may be necessary to account for a substantial population of hot Jupiters formed via this pathway.

\subsubsection{Initial System Configuration: Coplanar versus Inclined Perturbers } \label{sec:coplanar}

\begin{figure}[h]
    \centering
    \includegraphics[width=0.45\textwidth]{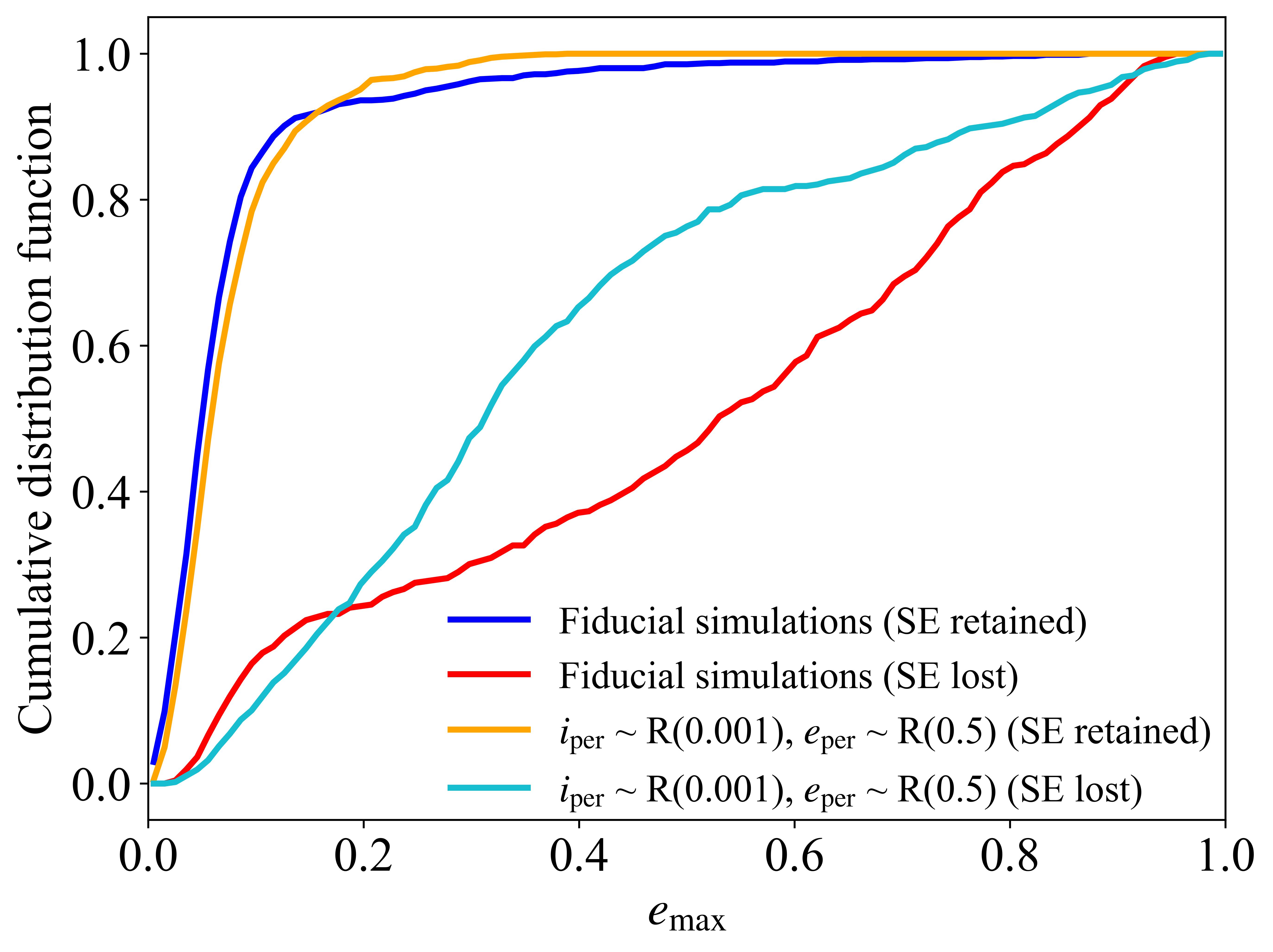}
    \caption{Cumulative distribution function of the maximum eccentricity reached by warm Jupiters. Blue and red curves show results from the fiducial simulation, while orange and cyan curves correspond to simulations with coplanar but eccentric perturbers, with initial eccentricities drawn from a Rayleigh distribution with $\sigma=0.5$.}
    \label{fig:high_e(SE&noSE)}
\end{figure}

As a robustness test of our main conclusion regrading eccentricity suppression by nearby super-Earths, we explore an alternative excitation channel in which the outer perturber is coplanar but highly eccentric, as proposed by \citet{Zink2023}. We conducted an additional set of simulations comprising 2000 systems. The key distinction in this test was the orbit of the outer perturber: it was initialized as coplanar (inclination $i_{\rm per}$ following a Rayleigh distribution with $\sigma=0.001$) yet highly eccentric (eccentricity $e_{\rm per}$ drawn from a Rayleigh distribution with $\sigma=0.5$). All remaining initial conditions replicated our fiducial setup detailed in Section \ref{sec:simulation setup}. The resulting eccentricity distributions are displayed in Figure \ref{fig:high_e(SE&noSE)}.

These simulations confirm that a coplanar, highly eccentric perturber can efficiently excite the eccentricity of a warm Jupiter in the absence of a nearby super‑Earth, although the efficiency of this excitation is lower than in our fiducial simulations. Particularly, only about $20\%$ of warm Jupiters reach maximum eccentricities $e_{\rm max} > 0.6$, a fraction significantly smaller than that produced in the fiducial case. In contrast, the presence of a nearby super‑Earth continues to provide strong damping, with most warm Jupiters maintaining low eccentricities, which is consistent with the high‑inclination case (Section \ref{sec:Nearby super-Earth}). The final eccentricity distribution of warm Jupiters depends on the initial eccentricity of the distant perturber, but a detailed characterization of this dependence is beyond the scope of the present work. Nevertheless, these results show that eccentricity suppression operates effectively whether eccentricity excitation is driven by vZLK oscillations induced by a highly inclined perturber or by secular perturbations from a coplanar but highly eccentric giant planet.

\subsection{Assessing the Relevance of Octupole-Level Timescales}\label{quad_oct}

In a subset of our simulations, eccentricity excitation of warm Jupiters is primarily driven by the vZLK mechanism induced by a mutually inclined distant perturber. In such configurations, the perturber's non-zero orbital eccentricity can introduce octupole-level secular effects, which are known to generate long-timescale modulations and, in some cases, extreme eccentricity growth beyond the quadrupole approximation \citep{Lithwick2011,Katz2011,Naoz2016,Weldon2024}. In this subsection, we assess whether our numerical setup and integration duration are sufficient to capture the relevant octupole-level dynamics in these vZLK-dominated systems. The relative strength of octupole to quadrupole contributions is measured by the octupole parameter \citep{Antognini2015}: 
\begin{equation}
\varepsilon_{\rm oct} \equiv \frac{a_{\mathrm{wj}}}{a_{\mathrm{per}}} \frac{e_{\mathrm{per}}}{1-e_{\mathrm{per}}^{2}}.
\end{equation} The characteristic timescale for octupole-driven evolution scales as $t_{\rm oct} \sim t_{\rm vZLK} / \sqrt{\varepsilon_{\rm oct}}$ up to factors of order unity depending on the exact definition adopted.

\begin{figure}[h]

    \centering
    \includegraphics[width=0.45\textwidth]{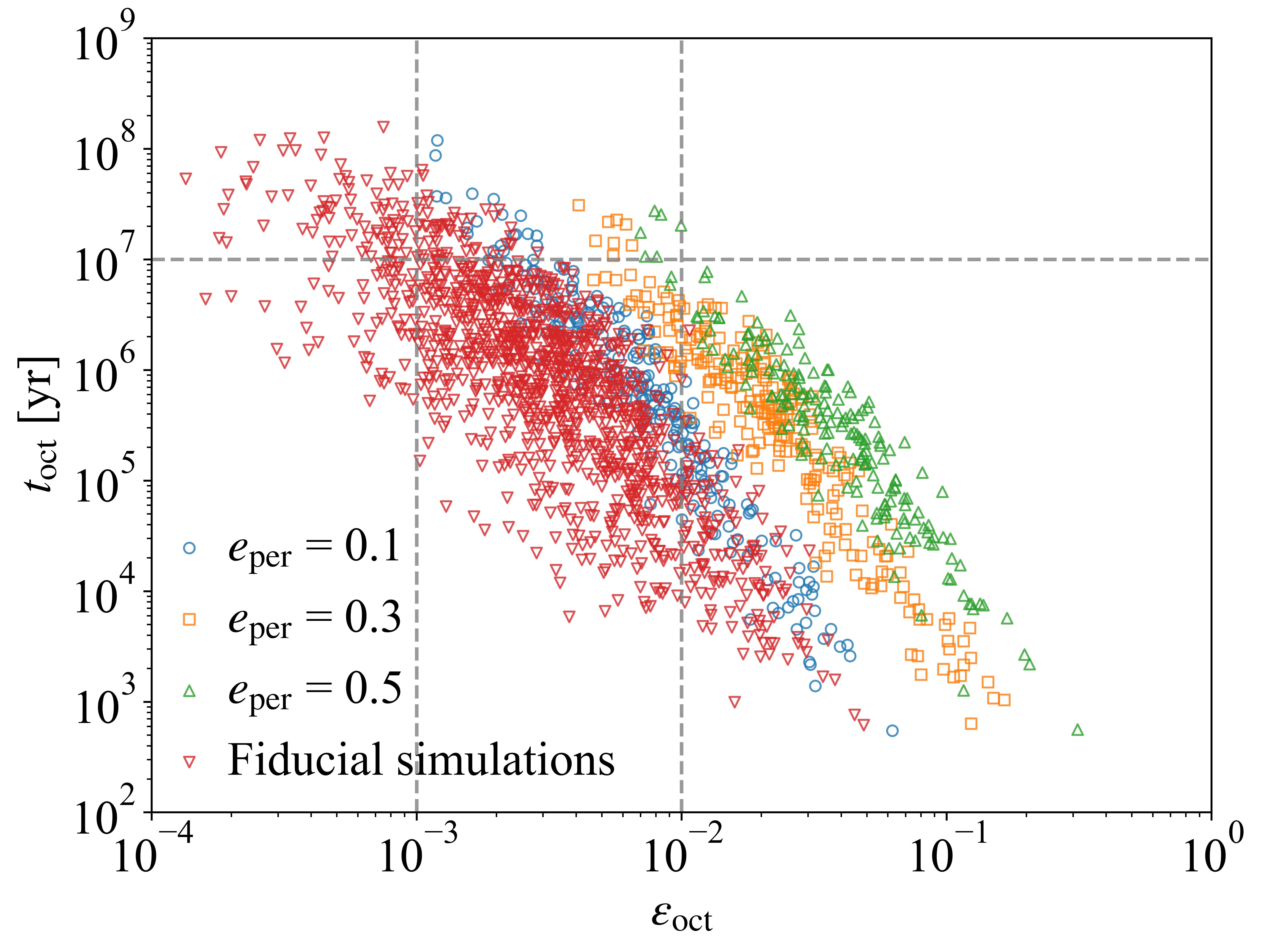}
    \caption{Joint distributions of the octupole strength parameter $\varepsilon_{\rm oct}$ and the corresponding octupole timescale $t_{\rm oct}$ for systems located in the ``vZLK window". Fiducial simulations and simulations with initial perturber eccentricities $e_{\rm per}=0.1,\,0.3,\,0.5$ are shown in red, blue, orange, and green, respectively. The dashed gray lines indicate $\varepsilon_{\rm oct}=0.01,\,0.001$ and $t_{\rm oct}=10^7$ yr.}
  \label{fig:epsilon_vs_toct}
\end{figure}

To focus on systems where vZLK oscillations are dynamically relevant, we restrict the following analysis to systems that lie within the “vZLK window”, defined here as those with an initial mutual inclination between the warm Jupiter and the perturber exceeding $40^\circ$, which approximately corresponds to the threshold for quadrupole‑order vZLK excitation when general relativistic precession is negligible. Figure \ref{fig:epsilon_vs_toct} shows the distribution of $\varepsilon_{\rm oct}$ and $t_{\rm oct}$ for this subset of systems. 

Our results indicate that systems with strong octupole coupling ($\varepsilon_{\rm oct} > 0.01$) typically have $t_{\rm oct}<10^7$ yr, implying that octupole-level effects would manifest within our integration time. Conversely, systems with $\varepsilon_{\rm oct}<0.001$ exhibit very weak octupole contributions, for which quadrupole-level dynamics dominate the secular evolution.  Importantly, the fraction of systems within the vZLK window that simultaneously satisfy $\varepsilon_{\rm oct} > 0.001$ and $t_{\rm oct}>10^7$ yr---i.e., systems for which octupole effects could be under sampled---is small: $2.8\%$ in our fiducial simulations, and $7.2\%$, $3.3\%$, $3.1\%$ for simulations with $e_{\rm per}=0.1,\,0.3,\,0.5$, respectively. 

Furthermore, within this small subset of potential concern, an additional suppression mechanism is at work. All warm Jupiters in these systems are initially accompanied by strongly coupled super-Earths (coupling parameter $\varepsilon > 100$). In these compact configurations, more than $90\%$ of the super-Earths survive the $10^7$-yr integration in our fiducial simulations and simulations with $e_{\rm per}=0.1$, while all super-Earths survive for $e_{\rm per}=0.3 $ and $e_{\rm per}=0.5$. This indicates that the vast majority of the inner multiplanet system remains stable and strongly coupled, effectively suppresses the excitation of the warm Jupiter's eccentricity via quadrupole-level vZLK mechanism, which in turn nullifies any octupole-level modulation dependent on it. As shown in Figure \ref{fig:epsilon_vs_toct}, even for the small set of simulations where warm Jupiters lose their nearby companions, these systems have $\varepsilon_{\rm oct} < 0.003$, placing them in the regime where octupole effects are inherently weak.

We therefore conclude that our adopted integration time of $10^7$ yr is sufficient to capture the relevant octupole-level secular dynamics in the vast majority of vZLK-active systems. Where octupole effects could be important (systems with large $\varepsilon_{\rm oct}$), the octupole timescale is well within our integration window, and where the octupole timescale is long, the octupole parameter is small enough that the quadrupole approximation remains valid. As a result, the eccentricity distributions reported in this work are not significantly affected by unaccounted octupole-level evolution.

\section{Comparison with observation}\label{sec:observation}

\subsection{Sample Selection}\label{sec:sample}

We compiled an initial sample of warm Jupiter systems with known eccentricities and masses from the Planetary Systems Composite Table \citep{PSCompPars,Akeson2013} \footnote{\url{https://exoplanetarchive.ipac.caltech.edu/}}, which aggregates data primarily from transit surveys and radial velocity (RV) surveys. For the observational sample, we define warm Jupiters as giant planets with orbital periods between 10 and 200 days and masses exceeding $0.25\,M_{\rm Jup}$. To expand the dataset,  we incorporated additional systems from dedicated studies of giant planets. To ensure consistency with the configuration used in Section \ref{sec:simulation setup}, we defined nearby companions as super-Earths or sub-Neptunes with orbital period ratios less than 4 relative to their warm Jupiters. We further excluded warm Jupiters accompanied by nearby giant planets, yielding an initial sample of 169 warm Jupiter systems. Because stellar companions are frequently observed around giant planets \citep{Ngo2015}, they can induce substantial changes in orbital architectures through secular perturbations \citep{Rice2022, Weldon2025}, which may mimic or even dominate the effects of planetary perturbers. To isolate the dynamical role of planetary companions, we used the census of known planet-hosting multiple systems compiled by \citet{Thebault2025}\footnote{The catalog is publicly available at: \url{https://lesia.obspm.fr/perso/philippe-thebault/plan_allbinall.txt} and \url{https://lesia.obspm.fr/perso/philippe-thebault/plan_circ.txt}.} and restricted our comparison sample to warm Jupiters with no detected stellar companions. The selection process is summarized as follows:

\begin{enumerate}
    \item[(1)] \textbf{Exclusion of nearby giant companions:} Removal of warm Jupiters with known giant‐planet companions within an orbital‐period ratio $< 4$ left 169 systems.

    \item[(2)] \textbf{Exclusion of stellar companions:} Using the catalog of \citet{Thebault2025}, we identified and excluded 39 systems with detected stellar companions. The remaining 130 warm Jupiter systems have no detected stellar companions.

    \item[(3)] \textbf{Subclassification by companion type and detection method:}
    \begin{itemize}
        \item \textbf{With nearby (super‐Earth/sub‐Neptune) companions:} 12 systems (eccentricities derived from RV or transit timing variations (TTVs)). These are listed for reference in Table \ref{Tab: WJ with nearby-com}.
        \item \textbf{Without nearby planetary companions:} 118 systems.
        \begin{itemize}
            \item \textbf{Eccentricities from RV: }109 systems. This forms the primary comparison sample for testing our simulations against observations (Section \ref{sec:obs} and \ref{sec:bias}).
            \item \textbf{Eccentricities from TTVs:} 9 systems.
        \end{itemize}
    \end{itemize}
\end{enumerate}

Thus, the final sample used for the statistical model comparison consists of 109 RV warm Jupiters without stellar or nearby planetary companions.

\setlength{\tabcolsep}{14pt}
\begin{table*}[t]
\centering
\caption{Warm Jupiters with nearby companions. The column ``Method" indicates the technique used to determine the orbital eccentricities of each warm Jupiter---either from RV measurements or TTVs.}
\begin{tabular}{cccccc}
    \toprule
    \\[-3.9ex]
    \toprule
    planet & $M_p[M_{\rm Jup}]$ & $P$ [day] & $e$ & Method & Ref. \\
    \midrule
    rho CrB b & $1.093^{+0.023}_{-0.023}$ & 39.84 & $0.038^{+0.0025}_{-0.0025}$ & RV & \citet{Brewer2023} \\
    HD 191939 e & $0.3530^{+0.0126}_{-0.0126}$ & 101.12 & $0.031^{+0.0008}_{-0.0016}$ & RV & \citet{Orell-Miquel2023} \\
    HIP 57274 c & $0.409^{+0.009}_{-0.009}$ & 32.03 & $0.05^{+0.02}_{-0.02}$ & RV & \citet{Fischer2012} \\
    Kepler-30 c & $2.01^{+0.16}_{-0.16}$ & 60.32 & $0.0111^{+0.001}_{-0.001}$ & TTVs & \citet{Sanchis-Ojeda2012} \\
    Kepler-56 c & $0.614^{+0.044}_{-0.044}$ & 21.41 & $0.0^{+0.01}_{-0.01}$ & RV & \citet{Otor2016} \\
    Kepler-87 b & $1.020^{+0.028}_{-0.028}$ & 114.73 & $0.036^{+0.009}_{-0.009}$ & TTVs & \citet{Ofir2014} \\
    Kepler-117 c & $1.84^{+0.18}_{-0.18}$ & 50.79 & $0.0323^{+0.0033}_{-0.0033}$ & TTVs & \citet{Bruno2015} \\
    Kepler-302 c & $2.936^{+1.658}_{-1.658}$ & 127.45 & $0.038^{+0.037}_{-0.037}$ & TTVs & \citet{Wu2018} \\
    KOI-142 c & $0.654^{+0.024}_{-0.024}$ & 22.27 & $0.0$ & RV & \citet{Weiss2024} \\
    TOI-216 c & $0.56^{+0.02}_{-0.02}$ & 34.55 & $0.0046^{+0.0027}_{-0.0012}$ & TTVs & \citet{Dawson2021} \\
    TOI-2525 c & $0.709^{+0.034}_{-0.034}$ & 49.26 & $0.152^{+0.006}_{-0.005}$ & RV & \citet{Trifonov2023} \\
    TOI-1670 c & $0.63^{+0.09}_{-0.08}$ & 40.75 & $0.09^{+0.05}_{-0.04}$ & RV & \citet{Tran2022} \\
    \bottomrule
\end{tabular}
\label{Tab: WJ with nearby-com}
\end{table*}

\begin{figure*}[t]
    \centering
    {
       \begin{minipage}[h]{0.45\linewidth}
       \includegraphics[width=\linewidth]{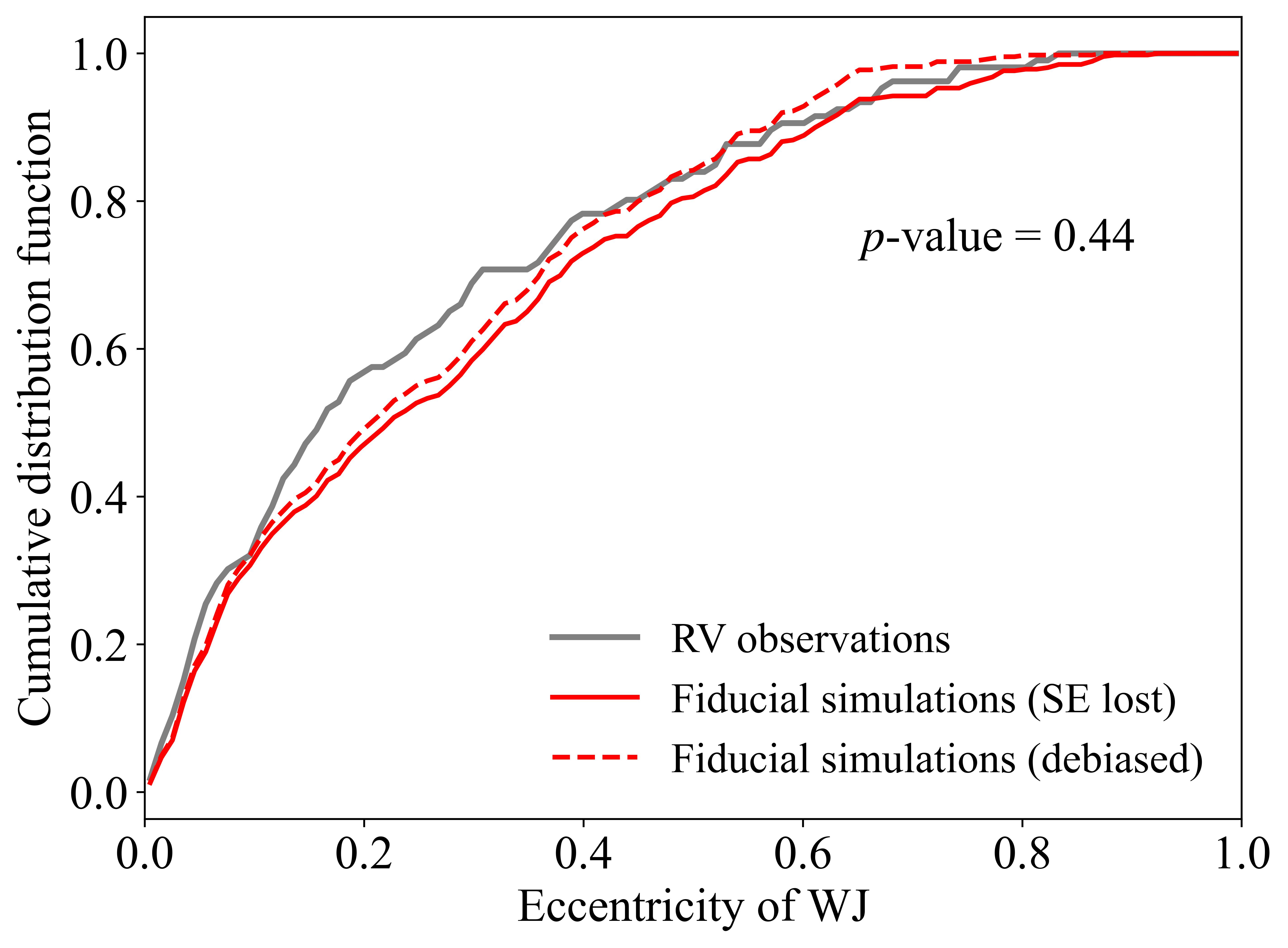}
       \end{minipage}
    }
    {
       \begin{minipage}[h]{0.45\linewidth}
       \includegraphics[width=\linewidth]{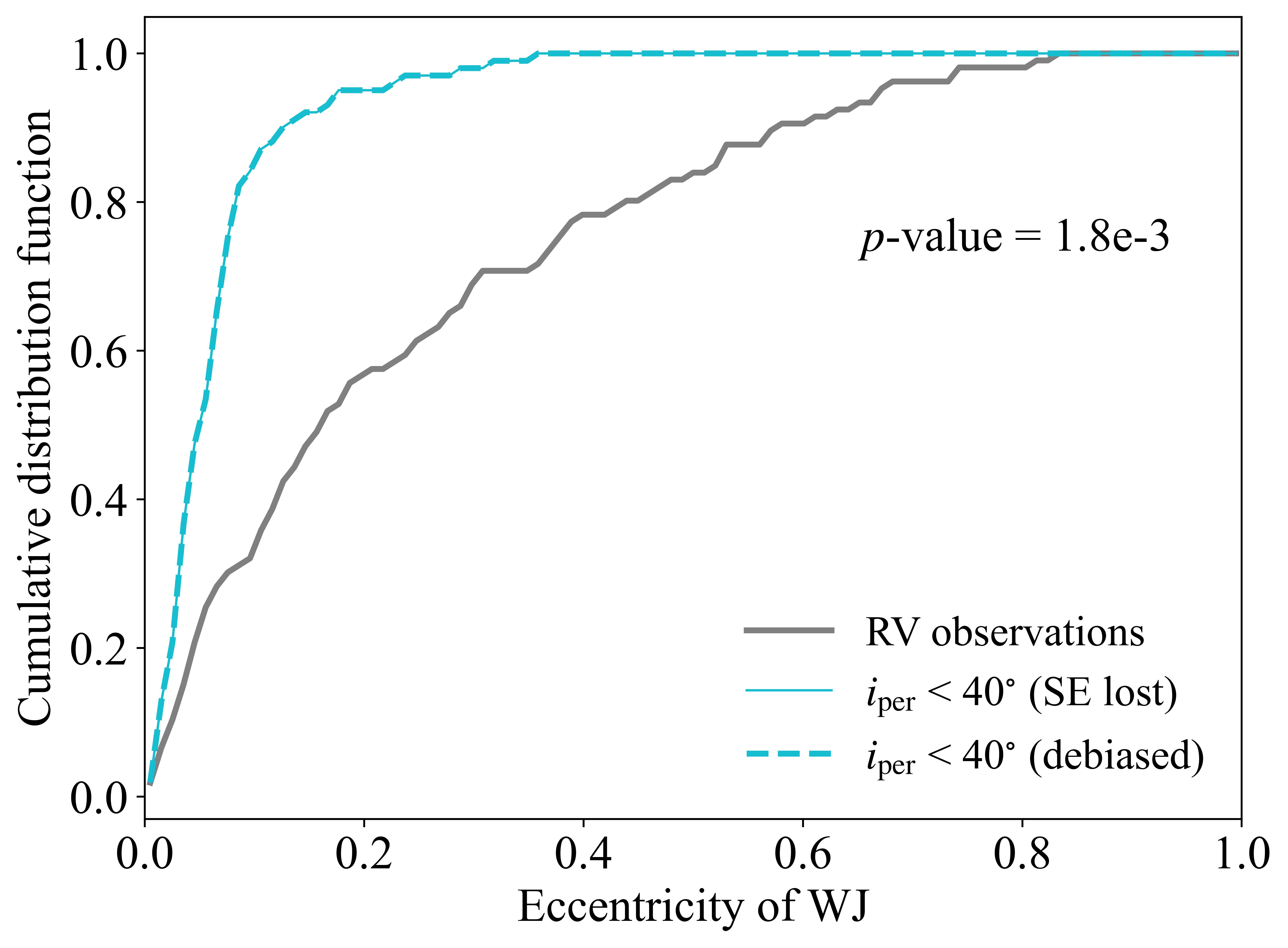}
       \end{minipage}
    }
    {
       \begin{minipage}[h]{0.45\linewidth}
       \includegraphics[width=\linewidth]{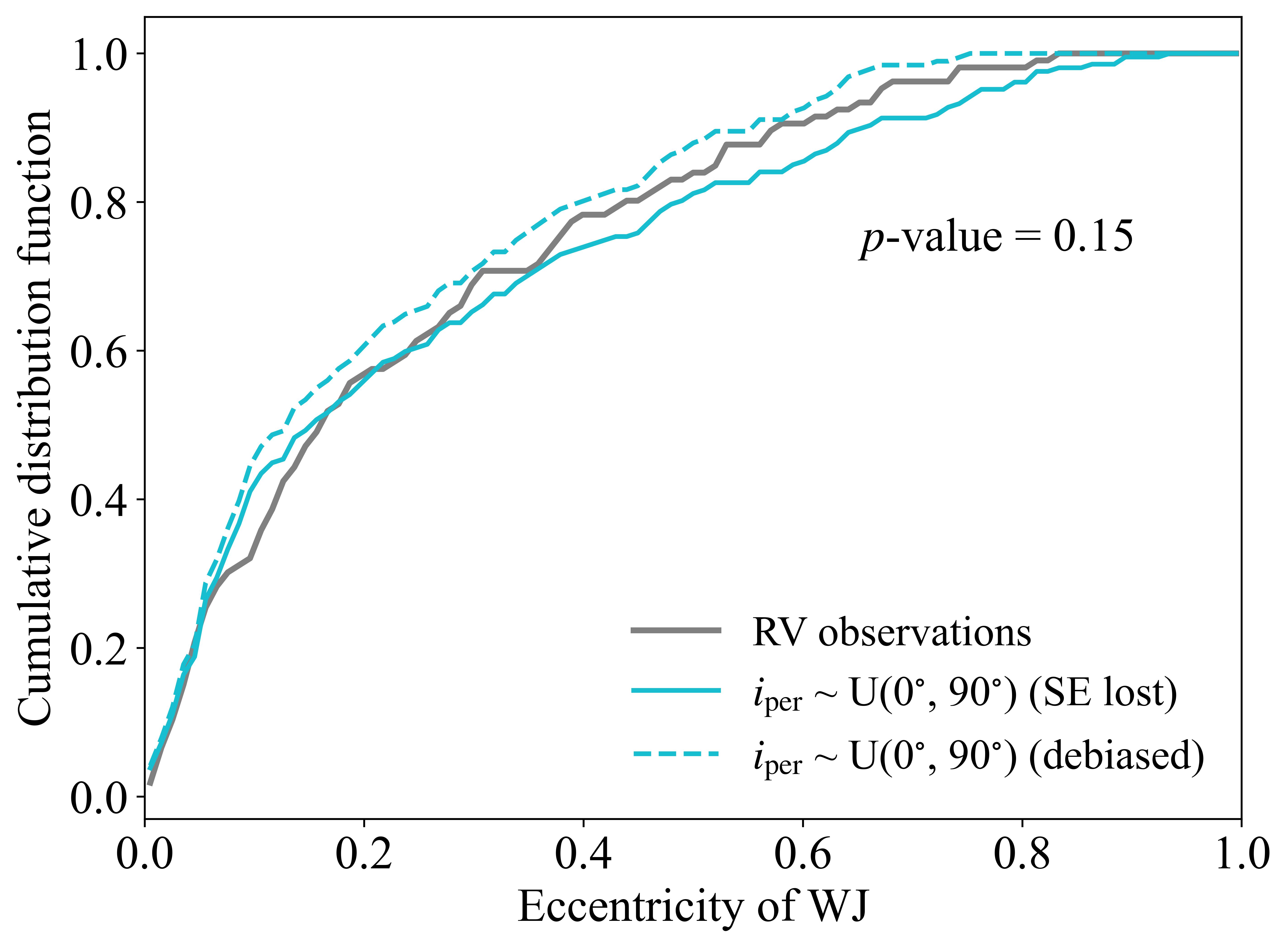}
       \end{minipage}
    }
    {
       \begin{minipage}[h]{0.45\linewidth}
       \includegraphics[width=\linewidth]{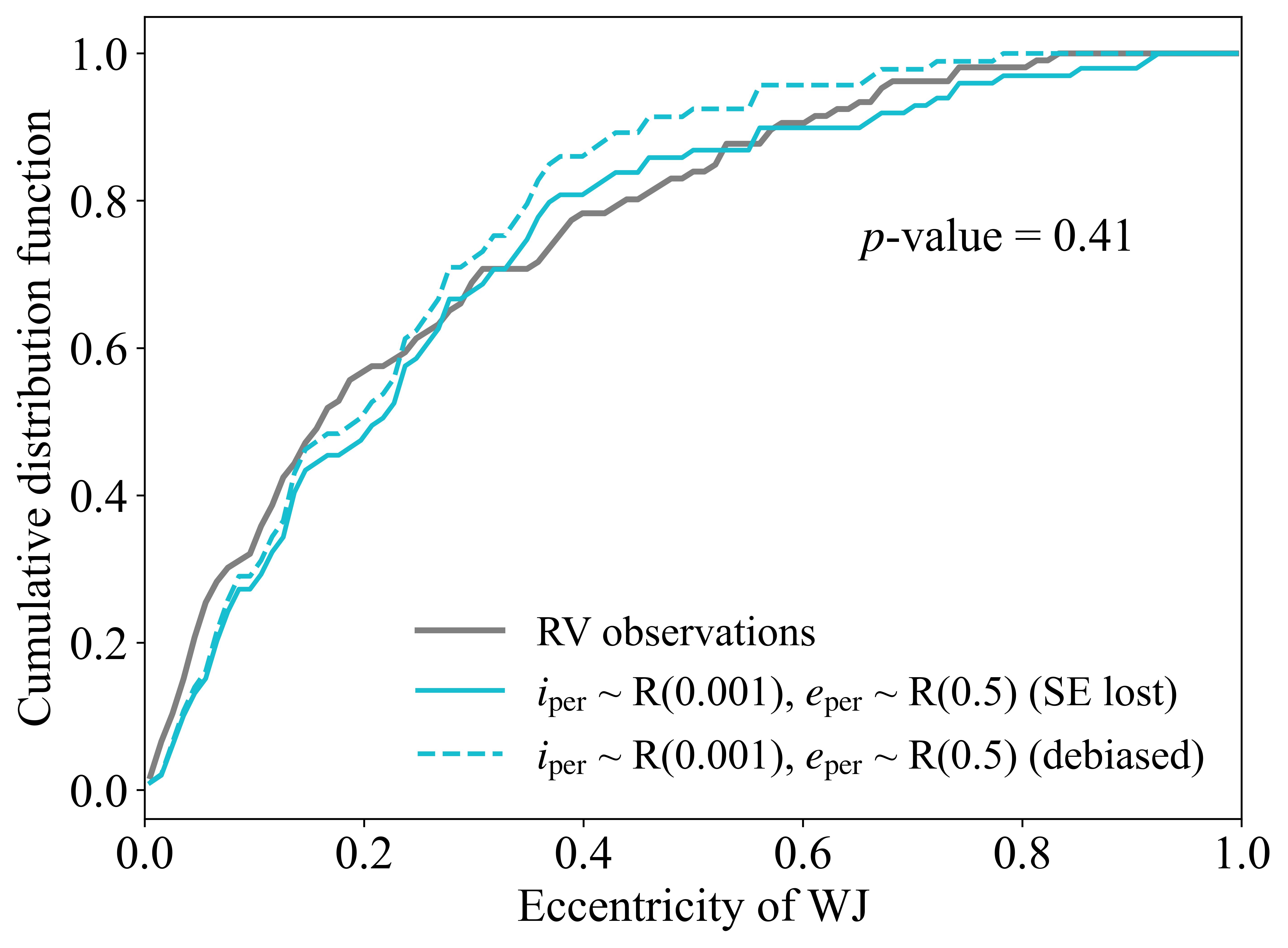}
       \end{minipage}
    }
    \caption{Comparison of the eccentricity distributions for warm Jupiters without nearby companions from our simulations (red and cyan lines) with RV observations (gray line). Simulations corrected for RV observational biases are shown as dashed curves. The $p$-values from the K-S tests between the bias-corrected simulations and RV observations are indicated in each panel. \textbf{Upper left}: Fiducial simulations. \textbf{Upper right}: Simulations with perturber inclinations $<40^\circ$ from the fiducial set. \textbf{Lower left}: Simulations with perturber inclinations uniformly distributed between $0^\circ$ and $90^\circ$. \textbf{Lower right}: Simulations with coplanar but eccentric perturbers ($\sigma=0.5$).}
    \label{fig:e(sim_obs)}
\end{figure*}

\subsection{Comparison Between Simulations and Observations}\label{sec:obs} 

Due to observational limitations and selection biases, the true maximum eccentricities of warm Jupiters are often unmeasurable. Therefore, we adopt the final eccentricities from our simulations for comparison with observations in this section. We first compare the simulated eccentricity distributions directly with observations, and subsequently examine the impact of observational selection effects.

In various simulations we have conducted, warm Jupiters that retain nearby companions typically exhibit orbital eccentricities around 0.05, consistent with the values reported in Table~\ref{Tab: WJ with nearby-com}. Therefore, in the following, we focus on eccentricity distribution comparison for  warm Jupiters without nearby companions. 

In our fiducial simulations, warm Jupiters that lose their nearby companions display final eccentricity distributions closely matching those of observed warm Jupiters without nearby companions, as identified via RV surveys (as shown in the upper left panel of Figure \ref{fig:e(sim_obs)}). A Kolmogorov–Smirnov (K–S) test yields a $p$-value of 0.19, indicating no statistically significant difference between the simulated and observed distributions. However, when restricting the analysis to fiducial simulations with low initial perturber inclinations ($i< 40^\circ$), the resulting eccentricity distribution of warm Jupiters without nearby companions fails to reproduce the observations (as shown in the upper right panel of Figure \ref{fig:e(sim_obs)}). 

To enable a population-level comparison, we performed an additional ensemble of 1000 simulations in which the perturber inclination is drawn from a uniform distribution between $0^\circ$ and $90^\circ$, while all other parameters are kept identical to those in the fiducial simulations (Section \ref{sec:simulation setup}). The resulting eccentricity distribution of warm Jupiters without nearby companions are shown in the lower left panel of Figure \ref{fig:e(sim_obs)}. A K–S test yields a $p$-value of 0.54 , indicating good agreement with the observed eccentricity distribution. We further examined the coplanar, high eccentricity perturber configuration described in Section \ref{sec:coplanar}, which yields a comparable $p$-value of 0.48.

Taken together, these results suggest that reproducing the observed eccentricity distribution of warm Jupiters without nearby companions generally requires the presence of dynamically “extreme” perturbers, with a substantial fraction of systems hosting either high-inclination companions or coplanar but highly eccentric ones. By contrast, the lower eccentricities observed in systems with nearby companions are not merely a consequence of dynamical stability requirements; rather, they reflect active suppression of secular eccentricity oscillations of warm Jupiters through gravitational coupling with nearby companions. 

\subsection{Influence of Observational Biases}\label{sec:bias}

We further examined how observational selection biases impact the detectability of planets with high eccentricities. To quantify the dependence of detection probability on eccentricity in RV surveys, we adopted the method proposed by \citet{Petrovich2016}. Specifically, we implemented a detection probability function characterized by a threshold eccentricity of $e_{\rm t} = 0.75$ and a standard deviation of $e_{\rm sd} = 0.15$, following the prescriptions of \citet{Cumming2004} and \citet{Petrovich2016}.

The de-biased eccentricity distributions of warm Jupiters that lose their nearby companions, derived from the simulations described above, are shown as dashed curves in Figure \ref{fig:e(sim_obs)}. For the low-inclination samples, few warm Jupiters reach high eccentricity; consequently, the eccentricity de-biasing procedure has negligible impact on these cases, and the K-S test continues to yield low $p$-values when comparing simulations with observations. After applying the de-biasing correction, the $p$-values for the fiducial simulations, the uniform-inclination ensemble, and the coplanar but eccentric perturber configuration become 0.44, 0.15, and 0.41, respectively. Overall, the de-biased simulations remain broadly consistent with the observed eccentricity distribution. 

The decrease in the $p$-value for the uniform-inclination case arises because the eccentricity correction shifts the simulated distribution toward lower eccentricities, thereby moving it away from the previously optimal agreement with the observations. In contrast, for the fiducial configurations, the $p$-values increases, as the de-biasing correction pulls the eccentricity distribution back from the high-eccentricity end toward values that more closely match the observed population.

In summary, reproducing the broad eccentricity distribution of warm Jupiters without nearby companions through gravitational perturbations from external bodies generally requires that a significant fraction of perturbers possess either substantial mutual inclinations or non-negligible orbital eccentricities.

We note that a recently published study by \citet{Morgan2025} presents an updated eccentricity distribution for a large sample of warm Jupiters derived primarily from radial velocity measurements. We have verified that the eccentricity distribution adopted in this work (warm Jupiters without detected stellar companions or nearby companions) is statistically indistinguishable from that reported by \citet{Morgan2025}, with a K-S test yielding a $p$-value of 0.98. This agreement further supports the robustness of our conclusions against the choice of observational sample.

\subsection{Influence of Stellar Companions on Warm Jupiter Eccentricities}

As discussed in Section \ref{sec:sample}, stellar companions are common among giant planet hosts and may act as external secular perturbers. For the RV warm Jupiter sample considered here (109 systems without detected stellar companions or nearby companions, and 33 systems with detected stellar companions but no nearby companions), however, we find no statistically significant difference between the eccentricity distributions of systems with detected stellar companions and those without. A K-S test yields a $p$-value of 0.67, indicating that the two samples are statistically indistinguishable. 

From a dynamical standpoint, distant perturbers—whether giant planets or stellar companions—can excite the eccentricities of warm Jupiters through similar secular mechanisms such as vZLK oscillations. Given the current observational limitations, including incomplete detection of close or low-mass stellar companions and the relatively small number of systems with confirmed stellar companions, our sample remains insufficient to robustly distinguish between planetary and stellar perturbers. As a result, any intrinsic differences between the two populations may be obscured in the present data.

\section{Conclusion}\label{sec:conclusion}

In this study, we investigate the dynamical evolution and eccentricity distribution of warm Jupiters in multiplanet systems, focusing on the influence of nearby super-Earth companions and distant massive perturbers. Our simulations reveal that strong dynamical coupling between warm Jupiters and nearby super-Earths enhances overall system stability. Moreover, the presence or absence of nearby super-Earths significantly shapes the orbital eccentricities of warm Jupiters, offering a potential explanation for the observed dichotomy in their eccentricity distributions. 

Specifically, we find that warm Jupiters accompanied by nearby super-Earths typically maintain low eccentricities ($\sim0.05$), a consequence of strong coupling that suppresses eccentricity excitation. In contrast, warm Jupiters without such nearby companions experience more substantial eccentricity oscillations driven by distant perturbers, leading to broader and higher eccentricity distributions. A key finding is that this damping effect is robust: it operates regardless of whether the excitation is driven by a highly inclined perturber or a perturber on a coplanar but highly eccentric orbit. This generality allows us to construct a unified framework. Within this framework, the observed eccentricity dichotomy of warm Jupiters arises from the combined effects of two conditions: (1) the presence or absence of a nearby super-Earth companion, and (2) the action of a dynamically effective external perturber on an extreme orbit, characterized by either high mutual inclination or high orbital eccentricity. Low-eccentricity warm Jupiters therefore predominantly reside in systems with nearby super-Earths, whereas high-eccentricity warm Jupiters are more likely to originate from systems lacking such companions but hosting an external perturber capable of driving strong secular excitation. 

Consistent with this framework, we find that simulated warm Jupiters that lose their nearby companions can reproduce the observed broad eccentricity distribution only under specific perturbing conditions. Systems in which the distant perturber remains confined to low inclinations and low eccentricities fail to generate sufficient eccentricity excitation. In contrast, good agreement with observations is obtained when the perturber population includes a significant subset of systems with either high mutual inclination or large orbital eccentricity.

Beyond shaping the eccentricity distribution of warm Jupiters, our simulations also indicate that a small subset of systems can be driven to very high eccentricities. In these cases, warm Jupiters may reach sufficiently small pericenter distances for tidal dissipation to become effective, potentially triggering inward tidal migration and subsequent evolution into hot Jupiters.

Despite these results, several limitations remain. First, our simulations adopt a simplified tidal migration threshold and do not include full tidal dissipation, which may overestimate the efficiency of inward migration. Second, our simulations do not explicitly consider long-term stellar perturbations. Although stellar and planetary companions could excite warm Jupiter eccentricities through similar secular mechanisms, stellar companions may more readily produce effective excitation owing to their typical masses and orbital configurations, potentially further shaping the eccentricity distribution of warm Jupiters. While our observational sample shows no statistically significant difference in eccentricity between systems with and without detected stellar companions, this null result may reflect observational incompleteness rather than the true absence of a stellar-companion effect. Finally, observational incompleteness---particularly for low-mass inner planets---may bias our comparisons. Future work should incorporate more realistic tidal physics, account for additional dynamical effects, and make use of data from \textit{TESS}, as well as forthcoming missions like \textit{PLATO} and China's \textit{ET} to further test and refine these predictions.

\section{Acknowledgments}
\label{section:acknowledgements}

We thank the anonymous referee for their insightful comments and constructive suggestions, which have significantly improved the scientific quality of this manuscript. We also thank Xiangning Su for helpful discussions. This work was supported by the National Natural Science Foundation of China (NSFC; grant Nos. 12573076, 12103003 and 11973094). Ying He acknowledges support from the Institute of Science and Technology for Deep Space Exploration (Nanjing University-Suzhou Campus) during the manuscript revision period. We also acknowledge the support from the Doctoral research start-up funding of Anhui Normal University. This research has made use of the NASA Exoplanet Archive, which is operated by the California Institute of Technology, under contract with the National Aeronautics and Space Administration under the Exoplanet Exploration Program.

\bibliography{bibliography}
\bibliographystyle{aasjournal}

\end{document}